\pdfoutput=1
\documentclass[%
preprint,
preprintnumbers,
nofootinbib,
amsmath,amssymb,
pra,
]{revtex4-2}

\usepackage{bm}
\usepackage{nicematrix}
\usepackage{upgreek}
\usepackage{xcolor}
\usepackage{IEEEtrantools}
\usepackage{dsfont}
\usepackage{siunitx}
\usepackage{mleftright}
\usepackage{mathtools}
\usepackage{booktabs}
\usepackage{multirow}
\usepackage{xspace}
\usepackage[caption=false]{subfig}
\usepackage{placeins}
\usepackage{orcidlink}

\usepackage{diffcoeff}
\difdef{l}{}{op-symbol = \mathrm{d}}

\PassOptionsToPackage{breaklinks}{hyperref}
\usepackage{hyperref}
\hypersetup{
	colorlinks = true,
	citecolor={blue},
	linkcolor={red},
	allbordercolors = {white},
	urlcolor={blue},
}
\usepackage{cleveref}

\newcommand{\upe}{\mathrm{e}}
\newcommand{\upi}{\mathrm{i}}

\newcommand{\ups}{\mathrm{s}}

\newcommand{\upT}{\mathrm{T}}
\newcommand{\upI}{\mathrm{I}}
\newcommand{\upII}{\mathrm{II}}

\newcommand{\Id}{\mathds{1}}
\newcommand{\avec}{\mathfrak{a}}
\newcommand{\PS}{R}
\newcommand{\BS}{B}

\newcommand{\vac}{\mathrm{vac}}

\newcommand{\Hc}{\mathrm{H.c.}}

\DeclareMathOperator{\diag}{diag}

\DeclareMathOperator{\Tr}{Tr}
\DeclareMathOperator{\Err}{Err}
\DeclareMathOperator{\rect}{rect}

\newcommand*{\mqty}[1]{
	\begin{pmatrix}
		#1
	\end{pmatrix}
}

\DeclarePairedDelimiter\abs{\lvert}{\rvert}
\DeclarePairedDelimiter\norm{\lVert}{\rVert}

\DeclarePairedDelimiter\ket{\lvert}{\rangle}
\DeclarePairedDelimiterX\innerp[2]{\langle}{\rangle}{#1\,\delimsize\vert\,\mathopen{}#2}
\DeclarePairedDelimiterX\outerp[2]{\lvert}{\rvert}{\,#1\delimsize\rangle\delimsize\langle\mathopen{}#2\,}
\DeclarePairedDelimiterX\braket[3]{\langle}{\rangle}{#1\,\delimsize\vert\,\mathopen{}#2\,\delimsize\vert\,\mathopen{}#3}

\newcommand*{\refcite}[1]{ref.~\onlinecite{#1}}

\newcommand{\IEEElabel}[1]{\addtocounter{equation}{-1}\refstepcounter{equation}\label{#1}}

\newcommand{\typezero}{type\protect\nobreakdash-0\xspace}

\newcommand{\typeI}{type\protect\nobreakdash-I\xspace}

\newcommand{\typezeroI}{type\protect\nobreakdash-0/I\xspace}

\newcommand{\typeII}{type\protect\nobreakdash-II\xspace}

\begin{document}
\count\footins = 1000

\title{Frequency-Resolved Simulations of Highly Entangled Biphoton States:\\Beyond the Single-Pair Approximation. I. Theory}

\author{\orcidlink{0009-0009-9005-702X}\,Philipp Kleinpa\ss}
\email{philipp.kleinpass@dlr.de}%
\affiliation{%
 Institute of Communications and Navigation, German Aerospace Center (DLR)
 }%
\affiliation{%
 Institute of Applied Physics, Technical University of Darmstadt%\\
}%

\author{\orcidlink{0000-0001-8114-1785}\,Thomas Walther and \orcidlink{0000-0002-1046-6176}\,Erik Fitzke}
\email{erik.fitzke@physik.tu-darmstadt.de}
\affiliation{%
 Institute of Applied Physics, Technical University of Darmstadt%\\
}%

\date{\today}% It is always \today, today,

\begin{abstract}
We discuss an expansion of the detection probabilities of biphoton states in terms of increasing orders of the joint spectral amplitude. The expansion enables efficient time- or frequency-resolved numerical simulations involving quantum states exhibiting a high degree of spectral entanglement. Contrary to usual approaches based on one- or two-pair approximations, we expand the expressions in terms corresponding to the amount of correlations between different pairs. The lowest expansion order corresponds to the limit of infinitely entangled states, where different pairs are completely uncorrelated and the full multi-pair statistics are inferred from a single pair. We show that even this limiting case always yields more accurate results than the single-pair approximation. Higher expansion orders describe deviations from the infinitely entangled case and introduce correlations between the photons of different pairs.
\end{abstract}

\maketitle

\section{Introduction}
Entanglement represents one of the most prominent examples of a purely quantum-mechanical phenomenon that shares no analogue with classical physics. Throughout the years, entangled states have evolved from a curiosity used to test the fundamental predictions of quantum physics~\cite{Bell_1964, Clauser_1969, Aspect_1982} to a valuable resource with many important applications, especially in quantum computation and quantum communication~\cite{Ekert_1991, BBM_1992, Bennett_1996, Tittel_2000}. In quantum optics, one of the most common sources of entanglement are biphoton states arising from the excitation of higher-order interaction processes such as spontaneous parametric down-conversion (SPDC) or spontaneous four-wave mixing (SFWM) within media exhibiting non-linear response functions~\cite{Grice_1997, Law_2000, Yang_2008, Helt_2010, Christ_2011, Euler_2021}.

Due to the probabilistic nature of these processes, whenever a pair of entangled photons is created, there is a non-vanishing probability of generating one or more additional pairs, an effect that may be detrimental to the performance of many applications employing such resources. To limit the impact of these multi-pair events, the intensity of the light used to pump the non-linear optical process can be lowered to a level where they become negligible. However, this also severely reduces the probability of generating any pair at all and, in consequence, results in a lower system performance.

The statistics of multi-pair generations are determined by the joint spectral amplitude (JSA) of the process. Nonetheless, many theoretical models do not consider the spectral degree of freedom~(DOF) at all or restrict their considerations to single-~\cite{Keller_1997, Grice_1997, Law_2000, Mikhailova_2008, Lee_2014, Liu_2020, Phehlukwayo_2020, Dorfman_2021} or two-pair~\cite{Ou_1999, Scarani_2005, Mueller_2020} generations. The latter require the evaluation of second and fourth order terms in the JSA, respectively, while the inclusion of higher-order pair generations leads to correspondingly more complex expressions.

A mathematical framework to describe the entire photon statistics, including all orders of multi-pair effects, is the phase-space formalism of Gaussian states~\cite{Ma_1990, Takeoka_2015, Fitzke_2023}. In this formalism, the biphoton spectrum can be included by representing the JSA in terms of its Schmidt decomposition~\cite{Schmidt_1907, Law_2000, Lamata_2005, Mauerer_2009}. In practice, a numerical Schmidt decomposition can be obtained by solving coupled integral equations~\cite{Schmidt_1907,Mauerer_2009_Thesis} or by performing either an orthogonal basis expansion~\cite{Lamata_2005} or a discretization of the spectrum~\cite{Law_2000,Mauerer_2009,Parker_2000} followed by a singular value decomposition of the resulting matrix. The latter approach was recently used in~\refcite{Thomas_2021}, where expressions for photon-number-resolved detection were derived and the effect of spectral and photon-number impurity on the Hong-Ou-Mandel interference visibility were examined.

For highly entangled biphoton states, however, implementing these methods directly may require considerable computational resources. The temporal and spectral correlations between the photons of an entangled pair are characterized by the contributing Schmidt modes and can be quantified by the Schmidt number~\cite{Fedorov_2006, Mikhailova_2008, Mauerer_2009, Horoshko_2018} of the JSA. A high degree of spectral entanglement is associated with a large Schmidt number and many contributing Schmidt modes. Therefore, numerically performing a Schmidt decomposition becomes computationally challenging for highly entangled biphoton states and infeasible when approaching the limit of infinitely strong entanglement.

A large Schmidt number is typically accompanied by the corresponding JSA exhibiting a large aspect ratio between the difference and the sum of the observable frequencies of the signal and idler photons, leading to an increasingly narrow grid required to perform a sufficiently accurate discretization of the frequency space. Ultimately, this also leads to the question of how the covariance formalism generalizes in the limit of a continuum of modes. Some mathematical properties of such continuous-mode Gaussian states have been examined in \refcite{Bhat_2019}. In \refcite{Nauth_2022}, the fact that the large aspect ratio limits the spread of higher orders of the JSA was used to enable a point-wise evaluation of the arising expressions.

In this work, we provide an alternative approach, intuitively exploiting the monogamy of entanglement, i.e.\ the fact that a larger amount of pairwise entanglement leads to a lower amount of correlations between distinct pairs~\cite{Coffman_2000, Osborne_2006}. For example, in the limit of an infinite amount of equally contributing Schmidt modes the pair statistics follow a Poisson distribution~\cite{Mauerer_2009}. Thus, in this case, the description of a single pair already contains all information about the entire multi-pair state. Within the framework of Gaussian states, we examine series expansions of the covariance matrix and the arising detection probabilities to obtain expressions in terms of a bivariate Poisson distribution, corresponding to infinitely strong entanglement and correction terms representing higher-order correlations. Such a description is convenient because, due to the weak correlations between different pairs, it allows to accurately describe higher numbers of generated pairs without the need of computing correspondingly higher orders of the JSA. For example, we show that the bivariate Poisson approximation only requires the evaluation of terms quadratic in the JSA and nonetheless is more accurate than the single-pair approximation in all parameter regimes. Similarly, including terms of fourth order in the JSA yields the lowest-order correction in terms of a bivariate Hermite distribution. To allow for a simple assessment of the applicability, we derive easy to evaluate bounds on the relative errors introduced by these expansions.

This work constitutes the first of a two-part series, focusing on the mathematical methods and their physical interpretation. In the second part of the series, \refcite{Kleinpass_2024_partII}, we demonstrate our methods by simulating entanglement-based quantum key distribution systems and validate compare the results to measurement data.

The remainder of this article is structured as follows. In \cref{sec:Gaussian_States} we introduce the formalism of Gaussian states and its generalization to a continuum of frequency modes. We present the series expansion of the arising detection probabilities in the general context of Gaussian states. In \cref{sec:HighlyEntangledStates} we review some important properties of entangled biphoton states before examining the series expansion of the renormalized covariance. To provide some further intuition to the physical interpretation of our approximations, the two lowest-order expansions in terms of the JSA are explicitly discussed in more detail.

\section{Gaussian States in Phase-Space}\label{sec:Gaussian_States}

We recap the well-known description of Gaussian states in \cref{sec:Discrete_modes_GS}, before discussing the extension of the formalism to the limit of a continuum of time and frequency modes in \cref{sec:Continuum_limit} and the inclusion of number-resolved detection in \cref{sec:Photon-Number-Resolved_Detection}. Lastly, in \cref{sec:Evaluating_the_Probabilities}, we discuss the evaluation of the detection probabilities in a continuous-mode setting.

\subsection{Discrete-Mode Gaussian States}
\label{sec:Discrete_modes_GS}

Vacuum, coherent, thermal and squeezed states, as well as biphoton states generated by SPDC or SFWM are prominent examples of so-called Gaussian quantum states $\hat{\varrho}$, defined by their characteristic function ${\mathcal{C}(\bm{\xi}) = \Tr\mleft( \hat{\varrho} \exp(\upi \bm{\xi}^\dag \hat{\bm{\avec}}) \mright)}$ being Gaussian~\cite{Wang_2007, Weedbrook_2012, Olivares_2012, Adesso_2014, Takeoka_2015}:
\begin{equation}
     \mathcal{C}(\bm{\xi}) = \exp\mleft( -\frac{1}{4} \bm{\xi}^\dag \bm{\gamma} \bm{\xi} + \upi \bm{\xi}^\dag \bm{\alpha} \mright) \,.
\end{equation}
The normal-mode vectors for a system with ${M \in \mathbb{N}}$ discrete DOFs read~\cite{Thomas_2021}
\begin{IEEEeqnarray}{C"t"C}\label{eq:NormalModeVector_Definition}
    \hat{\bm{\avec}} = \mqty{ \hat{\bm{a}}_1\\ \vdots\\ \hat{\bm{a}}_M\\ \hat{\bm{a}}_1^\dag\\ \vdots\\ \hat{\bm{a}}_M^\dag } & and &
    \hat{\bm{\avec}}_\dag = \mqty{ \hat{\bm{a}}_1^\dag\\ \vdots\\ \hat{\bm{a}}_M^\dag\\ \hat{\bm{a}}_1\\ \vdots\\ \hat{\bm{a}}_M }\,.
\end{IEEEeqnarray}
In general, within each discrete DOF, some additional continuous DOF can be specified, such as time, space, frequency or momentum. Discretizing these quantities on a sufficiently fine grid results in an additional ${N \in \mathbb{N}}$~modes within each of the~$M$ discrete DOFs. Therefore, the elements of $\hat{\bm{\avec}}$ in \cref{eq:NormalModeVector_Definition} are vectors themselves, with their components given by the creation and annihilation operators of the corresponding modes:
\begin{subequations}
    \begin{IEEEeqnarray}{rCr?l?l?l}
        \hat{\bm{a}}_k &=& \Big( \hat{a}_{k,\omega_1} & \hat{a}_{k,\omega_2} & \dots & \hat{a}_{k,\omega_N} \Big)^\upT \,, \IEEEeqnarraynumspace\\
        \hat{\bm{a}}_k^\dag &=& \Big( \hat{a}_{k,\omega_1}^\dag & \hat{a}_{k,\omega_2}^\dag & \dots & \hat{a}_{k,\omega_N}^\dag \Big)^\upT \,.
        \IEEEeqnarraynumspace
    \end{IEEEeqnarray}    
\end{subequations}
Here, ${k = 1, \dots, M}$ labels the discrete DOFs and $\omega_l$ with ${l = 1, \dots, N}$ labels the discretized continuous DOFs, which we take to be different frequency components.\\
Gaussian states are fully characterized by their displacement vector ${\bm{\alpha} = \langle \hat{\bm{\avec}} \rangle}$ and positive definite covariance matrix
\begin{equation}\label{eq:CovarianceDisplacement_BlockStructure}
    \bm{\gamma} = 
    \mqty{
        \bm{A} & \bm{B}\\
        \bm{B}^\dag & \bm{A}^*
    }  = \mqty{\bm{\gamma}_{1\,1} & \dots & \bm{\gamma}_{1\,2M}\\ \vdots & \ddots & \vdots\\ \bm{\gamma}_{2M\,1} & \dots & \bm{\gamma}_{2M\,2M}}\,,
\end{equation}
where $\bm{A}$ and $\bm{B}$ are ${M \times M}$ block matrices representing the $M$ discrete DOFs and each block
\begin{equation}\label{eq:CovarianceDisplacement_Definition}
    \bm{\gamma}_{kl} = \big\langle  \hat{\bm{\avec}} \hat{\bm{\avec}}_\dag^\upT \big\rangle_{kl} + \big\langle \hat{\bm{\avec}}_\dag \hat{\bm{\avec}}^\upT \big\rangle_{lk} - 2 \big\langle \hat{\bm{\avec}} \big\rangle_k \langle \hat{{\bm{\avec}}}_\dag^\upT \big\rangle_l
\end{equation}
of the covariance is an ${N \times N}$ matrix representing the discretization of the continuous DOFs.\nolinebreak\footnote{Other formulations of the covariance exist in the literature, differing by a factor of two~\cite{Olivares_2012,Hamilton_2017,Kruse_2019} or defining a real-valued covariance with respect to the quadrature basis~\cite{Olivares_2012,Adesso_2014,Wang_2007}.} An example of a Gaussian state is the vacuum state, characterized by ${\bm{\gamma} = \mathds{1}}$ and ${\bm{\alpha} = \bm{0}}$, where $\mathds{1}$ is the identity operator.

Transformations described by Hamiltonians linear or quadratic in the creation and annihilation operators map Gaussian states to other Gaussian states. While linear interactions simply correspond to translations of the displacement vector, quadratic interactions can always be written in the form~\cite{Adesso_2014, Horoshko_2018, Thomas_2021}
\begin{IEEEeqnarray}{c"c"c}\label{eq:QuadraticUnitaryTransformation_Definition}
    \hat{U} = \upe^{-\upi \hat{H}} \,,&
    \text{where} &
    \hat{H} = \frac{1}{2} \hat{\bm{\avec}}^\dag \bm{H} \hat{\bm{\avec}} \,,
    \IEEEeqnarraynumspace
\end{IEEEeqnarray}
with a Hermitian matrix $\bm{H}$. The state's evolution caused by such interactions is obtained by transforming the covariance matrix and displacement vector according to~\cite{Adesso_2014}
\begin{IEEEeqnarray}{c"c"c}\label{eq:CovarianceDisplacement_SymplecticTransformation}
    \bm{\gamma} \to \bm{S} \bm{\gamma} \bm{S}^\dag & \text{and} &
    \bm{\alpha} \to \bm{S} \bm{\alpha} \,,
    \IEEEeqnarraynumspace
\end{IEEEeqnarray}
with the complex-valued transformation matrix~\cite{Arvind_1995, Adesso_2014}
\begin{IEEEeqnarray}{c"c"c}\label{eq:SymplecticTransformationFromHamiltonian}
    \bm{S} = \upe^{\bm{Z}} \,, &
    \text{where} &
    \bm Z = -\upi \mqty{ \mathds{1} & 0\\ 0 & -\mathds{1} } \bm{H} \,,
    \IEEEeqnarraynumspace
\end{IEEEeqnarray}
fulfilling the symplectic relation
\begin{equation}
    \bm{S} \mqty{ \mathds{1} & 0\\ 0 & -\mathds{1} } \bm{S}^\dag = \mqty{ \mathds{1} & 0\\ 0 & -\mathds{1} }\,.
\end{equation}
Thus, the matrix $\bm{S}$ is directly obtained by writing the unitary transformation in the form of \cref{eq:QuadraticUnitaryTransformation_Definition}. An important subgroup of symplectic transformations are so-called passive transformations, corresponding to unitary transformation matrices which, in contrast to active transformations, preserve the state's photon number~\cite{Arvind_1995, Adesso_2014}.\\
Many quantum-optical setups employ single-photon detectors limited to distinguish between the presence and absence of vacuum, with the probability of obtaining a vacuum detection result given by~\cite{Takeoka_2015, Thomas_2021}
\begin{equation}\label{eq:DetectionProbability_Takeoka}
    P_{\vac} = 2^{MN} \frac{\exp\mleft( -\bm{\alpha}^\dag (\mathds{1} + \bm{\gamma})^{-1} \bm{\alpha} \mright)}{\sqrt{\det\mleft( \Id + \bm{\gamma} \mright)}} \,.
\end{equation}
This expression is invariant under permutations of the elements in the normal-mode vector in \cref{eq:NormalModeVector_Definition}. Therefore, the elements of the displacement vector and the covariance can always be reordered by simultaneously swapping the corresponding row-column-pairs.\\
A detection over a subset of the modes can be considered by applying an orthogonal projection $\bm{P}$ to the covariance and displacement:
\begin{IEEEeqnarray}{c"c"c"c}\label{eq:Projection}
    \bm{\gamma} \to \bm{P} \bm{\gamma} \bm{P} \,, &
    \bm{\alpha} \to \bm{P} \bm{\alpha} \,, &
    \text{where} &
    \bm{P} = \bm{P}^2 \,.
    \IEEEeqnarraynumspace
\end{IEEEeqnarray}
Such projections correspond to taking the partial trace over the state's density operator to obtain the remaining subsystem and preserve the Gaussian state property~\cite{Adesso_2014}. In \cref{sec:Appendix_Gaussian_Transformations}, we present all Gaussian transformations relevant for this work as well as their continuous-mode counterparts.

\subsection{The Continuous-Mode Limit}
\label{sec:Continuum_limit}
Most commonly, the Gaussian state formalism is used to describe the transformations of a finite number of discrete modes. In the majority of quantum optical systems, where the states are not localized within some kind of cavity, a continuum of time and frequency modes is present. This section describes how the formalism can be extended to account for such continuous degrees of freedom.

To allow for a description of states featuring an infinite amount, or even a continuum of modes, we rewrite \cref{eq:DetectionProbability_Takeoka} in the form
\begin{equation}\label{eq:DetectionProbability_RenormalizedCovariance}
    P_{\vac} = \frac{\exp\mleft( -\bm{\alpha}^\dag (\mathds{1} + \bm{\varGamma})^{-1} \bm{\alpha}/2 \mright)}{\sqrt{\det(\mathds{1} + \bm{\varGamma})}} \,,
\end{equation}
where
\begin{equation}\label{eq:RenormalizedCovariance_Definition}
    \bm{\varGamma} = \frac{\bm{\gamma} - \Id}{2}
\end{equation}
was introduced. We will refer to $\bm{\varGamma}$ as the \textit{renormalized covariance}, as it removes the (infinite) vacuum contributions from the covariance. In the continuous-mode limit, the renormalized covariance preserves the structure of \cref{eq:CovarianceDisplacement_BlockStructure}, i.e.\ it is still a ${2M \times 2M}$ matrix, however its blocks $\bm{\varGamma}_{kl}$ become integral operators with the corresponding kernels given by\nolinebreak\footnote{The generalization is straight forward: In the discretized case, the elements $\bm{\varGamma}_{kl}$ of the renormalized covariance are ${N \times N}$ matrices acting on some vector $\bm{x}_l$ via matrix multiplication: ${(\bm{\varGamma}_{kl} \bm{x}_l)_m = \sum_{n=1}^N (\bm{\varGamma}_{kl})_{mn} (\bm{x}_l)_n}$. In the continuous-mode limit ${N \to \infty}$, the sum is replaced by an integral and the discrete index $m$ becomes a continuous variable $\omega$. Thus we have ${(\bm{\varGamma}_{kl} \bm{x}_l)(\omega) = \int \dl\omega' \varGamma_{kl}(\omega,\omega') {x}_l(\omega')}$, where integrals without bounds are taken over the whole space of interest. The action of the total renormalized covariance $\bm{\varGamma}$ on some vector ${\bm{x} = (\bm{x}_1, \bm{x}_2, \dots, \bm{x}_{2M})}$ is given by ${(\bm{\varGamma} \bm{x})_k(\omega) = \sum_{l=1}^{2M} (\bm{\varGamma}_{kl} \bm{x}_l)(\omega)}$.}
\begin{equation}
    \varGamma_{kl}(\omega,\omega') = \frac{1}{2} \big\langle \hat{\avec}_k(\omega) \hat{\avec}_l^\dag(\omega') + \hat{\avec}_l^\dag(\omega') \hat{\avec}_k(\omega) \big\rangle - \big\langle \hat{\avec}_k(\omega) \big\rangle \big\langle \hat{\avec}_l^\dag(\omega') \big\rangle - \delta_{kl} \delta(\omega-\omega')  \,.
\end{equation}
One of the most important properties of the renormalized covariance $\bm{\varGamma}$ is that it is a trace class operator~\cite{Gohberg_2000}. An intuitive reason for this is given by the fact that ${\int \dl\omega |\varGamma_{mm}(\omega,\omega)| < \infty}$ for all ${m \leq M}$,\nolinebreak\footnote{Strictly speaking, this is not a sufficient condition but may still serve as valuable intuition according to \refcite{Simon_2005}: \emph{"If an integral operator with kernel K occurs in
some “natural” way and ${\int \abs{K(x, x)} \dl x < \infty}$, then the operator can (almost always) be proven to be trace class"}. A sufficient condition is given e.g. by $K$ being a Schwartz function~\cite{Brislawn_1988, Bornemann_2010}.} which is satisfied due to the direct correspondence between the trace of the renormalized covariance and the expectation value of the photon number operator:
\begin{equation}\label{eq:Covariance_Energy}
    \langle\hat{n}\rangle =  \frac{1}{2} \sum_{m=1}^{2M} \int \dl\omega \mleft( \varGamma_{mm}(\omega,\omega) + \big\langle \hat{\avec}_j(\omega) \big\rangle \big\langle \hat{\avec}_j^\dag(\omega) \big\rangle \mright) = \frac{\Tr (\bm{\varGamma}) + \bm{\alpha}^\dag \bm{\alpha}}{2}  < \infty \,.
\end{equation}
A more rigorous discussion about the properties of Gaussian states featuring an infinite amount of modes can be found in \refcite{Bhat_2019}, where $\bm{\varGamma}$ being trace class is introduced as a necessary condition for the corresponding covariance to describe a Gaussian quantum state. The renormalized covariance being trace class is significant as it allows for a generalization of the determinant in \cref{eq:DetectionProbability_RenormalizedCovariance} to a 
Fredholm determinant~\cite{Fredholm_1903, Bornemann_2010}, fulfilling
\begin{equation}\label{eq:Fredholm_Determinant}
    \det(\mathds{1} + \bm{\varGamma}) = \exp\{\Tr[\ln(\mathds{1} + \bm{\varGamma})]\}
\end{equation}
and ensuring that \cref{eq:DetectionProbability_RenormalizedCovariance} is well-defined in the continuous-mode limit.

\subsection{Photon-Number-Resolved Detection}\label{sec:Photon-Number-Resolved_Detection}

With recent advances in developments and applications of photon-number resolved detection~\cite{Cahall_2017,He_2022,Cheng_2023} and photon-number resolving detectors becoming commercially available~\cite{id281_brochure}, the inclusion of the full counting statistics in theoretical models has become an increasingly relevant topic. As the covariance formalism naturally contains the complete information about the photon statistics of the Gaussian state, it is well suited to model photon-number-resolved detection, e.g.\ by calculating the photon statistics from the Hafnian and loop Hafnian function of matrices derived from the covariance matrix~\cite{Quesada_2019_Simulating,Quesada_2019_FranckCondon,Bjoerklund_2019,Quesada_2022_Quadratic,Quesada_2018,Bulmer_2022_Threshold}.

Another approach of obtaining the photon-number distribution is to introduce the matrix ${\bm W = \diag(\bm w)^{\oplus 2}}$ into \cref{eq:DetectionProbability_RenormalizedCovariance}, where $\bm{X}^{\oplus 2} = \bm{X} \oplus \bm{X}$ indicates the direct sum of an operator $\bm{X}$ with itself. This generalizes the vacuum probability to a probability-generating function for the photon statistics~\cite{Fitzke_2023}:
\begin{equation}\label{eq:GeneratingFunctionGaussian}
    G(\bm{w}) = \frac{\exp\mleft( -\bm{\alpha}^\dag (\mathds{1} + \bm{W} \bm{\varGamma})^{-1} \bm{\alpha}/2 \mright)}{\sqrt{\det(\mathds{1} + \bm{W} \bm{\varGamma})}} \,.
\end{equation}
For $D$ detectors, the $M_d$ discrete modes sent into detector ${d = 1, \dots, D}$ are labeled as ${m_d = 1_d, \dots, M_d}$, with the total number of discrete modes ${M = \sum_d M_d}$. This introduces additional structure to the renormalized covariance, as we can write
\begin{IEEEeqnarray}{c}
    \bm{A} =
    \begin{pNiceArray}[last-col]{ccc}
        \bm{A}_{11} & \dots & \bm{A}_{1D} & \Block{3-1}{\mleft.\rule{0em}{35pt}\mright\}\substack{\text{$D$ detectors}}}\\
        \vdots& \ddots & \vdots\\
        \bm{A}_{D1} & \dots & \bm{A}_{DD}\\
        \CodeAfter
            \UnderBrace[shorten, yshift=5pt]{3-1}{3-3}{\substack{\text{$D$ detectors}}}
    \end{pNiceArray} \,, \IEEEnonumber\\
\end{IEEEeqnarray}
corresponding to the different detector combinations and
\begin{IEEEeqnarray}{c}
    \bm{A}_{jk} = 
    \begin{pNiceArray}[last-col]{ccc}
        \bm{A}_{1_j1_k}(\omega, \omega') & \dots & \bm{A}_{1_jM_k}(\omega, \omega') & \Block{3-1}{\mleft.\rule{0em}{35pt}\mright\}\substack{\text{$M_j$ discrete}\\\text{\vphantom{$M_j$}modes in}\\\text{detector $j$}}}\\
        \vdots & \ddots & \vdots &\\
        \bm{A}_{M_j1_k}(\omega, \omega') & \dots & \bm{A}_{M_jM_k}(\omega, \omega') &\\
        \CodeAfter
        \UnderBrace[shorten, yshift=5pt]{3-1}{3-3}{\substack{\text{$M_k$ discrete modes in detector $k$}}}
    \end{pNiceArray} \IEEEnonumber\\
\end{IEEEeqnarray}
representing the modes within each detector. The same applies for $\bm{B}_{jk}$ in \cref{eq:CovarianceDisplacement_BlockStructure}. With ${\bm{w} = (w_1, \dots, w_D)}$, the multivariate probability distribution for the detection of ${\bm{n} = (n_1, n_2, \dots n_D)}$ photons can be calculated by repeated differentiation~\cite{Thomas_2021, Nauth_2022, Fitzke_2023}:
\begin{equation}\label{eq:PND_Formuala}
    P(\bm n) = \bm{\mathcal{D}_{\bm{w}}^{\bm{n}}}\, G(\bm{w}) = \mleft(\prod_{d=1}^D\frac{1}{n_d!} \diffp[n_d]{}{-w_d}\mright) G(\bm{w}) \bigg\vert_{\bm{w}=\bm{1}} \,.
\end{equation}
Replacing $\bm{w}$ by simple functions of the differentiation parameters yields expressions for the generating functions of the moments and factorial moments of the photon statistics~\cite{Fitzke_2023}.

When the Hafnian-based approach is used to calculate the photon number distribution, the size of the matrix of which the Hafnian is calculated scales with the number of modes. All modes are considered independently, leading to discrete spatial modes and frequency modes being treated equally. In contrast, the generating-function-based approach is well suited to model the simultaneous detection of multiple modes because the number of required derivatives only depends on the number of detectors and photons instead of the total number of modes~\cite{Thomas_2021}. This advantage of the generating-function approach allows for a natural generalization of the expressions for the photon statistics to continuous-mode Gaussian states without adding complexity to the computation of the derivatives. For the Hafnian-based approach, a generalization to a continuum of modes is yet to be developed.

\subsection{Retrieval of the Detection  Probabilities}\label{sec:Evaluating_the_Probabilities}
To obtain the detection probabilities, the determinant in \cref{eq:GeneratingFunctionGaussian} needs to be evaluated. A direct approach would be to compute all non-zero eigenvalues of $\bm{\varGamma}$, which, however, quickly becomes numerically infeasible for kernels with narrow diagonal and anti-diagonal shapes arising for highly entangled biphoton states. Alternative methods involving the discretization on a rectilinear grid and applying quadrature rules to evaluate the determinant numerically have been discussed in the literature~\cite{Bornemann_2010} but suffer from the same limitation: the large amount of discretization points required renders these methods impractical for highly entangled states. 

\subsubsection{Series Expansion of the Determinant}
Especially in the regime of small mean photon numbers, the eigenvalues of the renormalized covariance are sufficiently close to zero for the Neumann series and the Taylor expansion of the logarithm to converge.\nolinebreak\footnote{For the series to converge the condition ${|\Lambda_1'| < 1}$ needs to be fulfilled, where $|\Lambda_1'|$ refers to the eigenvalue with the largest modulus of the covariance after applying all transformations.} Therefore, the Neumann series expansion of the inverse~\cite{Werner_2006},
\begin{equation}
        (\mathds{1} + \bm{W} \bm{\varGamma})^{-1} = \sum_{n=0}^\infty (-\bm{W} \bm{\varGamma})^n \,,
\end{equation}
and the Taylor expansion of the logarithm in \cref{eq:Fredholm_Determinant},
\begin{equation}\label{eq:FredholmDeterminant_LogarithmExpansion}
 \det\mleft( \Id + \bm{W} \bm{\varGamma} \mright) = \exp\mleft( -\sum_{n=1}^\infty \frac{(-1)^n}{n} \Tr\mleft[ (\bm{W} \bm{\varGamma})^n \mright] \mright)\,,
\end{equation}
can be used to rewrite \cref{eq:GeneratingFunctionGaussian} as
\begin{equation}
    G(\bm{w}) = \exp\mleft( \sum_{n=1}^\infty \frac{(-1)^n}{2} \mleft( \bm{\alpha}^\dag (\bm{W}\bm{\varGamma})^{n-1}\bm{W}
    \bm{\alpha} + \frac{\Tr\mleft[ (\bm{W} \bm{\varGamma})^n \mright]}{n} \mright) \mright) \,.
\end{equation}
Using \cref{eq:Covariance_Energy}, the lowest order of this expansion corresponds to independent Poissonian statistics:
\begin{equation}
    P(\bm{n}) \approx  \bm{\mathcal{D}}_{\bm{w}}^{\bm{n}} \, \upe^{-[\bm{\alpha}^\dag \bm{W} \bm{\alpha} + \Tr(\bm{W} \bm{\varGamma})]/2} 
    = \prod_{d=1}^D \frac{\langle \hat{n}_d \rangle^{n_d}}{n_d!}\, \upe^{-\langle \hat{n}_d \rangle} \,.
\end{equation}
Higher expansion orders represent corrections to these statistics, which immediately shows that at least one more term in the expansion must be included to account for correlations between different detectors, as they are expected for entangled biphoton states.

For the remainder of this work, we will consider states with ${\bm{\alpha} = \bm{0}}$. In this case, the second-order expansion reads
\begin{equation}
    G(\bm{w}) \approx \exp\mleft( -\sum_{d=1}^D w_d \langle \hat{n}_d \rangle + \sum_{d,d'=1}^D w_d w_{d'} \frac{\norm{\bm{A}_{dd'}}_{\mathrm{HS}}^2 + \norm{\bm{B}_{dd'}}_{\mathrm{HS}}^2}{2} \mright)
\end{equation}
and introduces correlations in terms of the Hilbert-Schmidt norm $\norm{\cdot}_{\mathrm{HS}}$ of the renormalized covariance.

\begin{figure}
    \centering
    \includegraphics{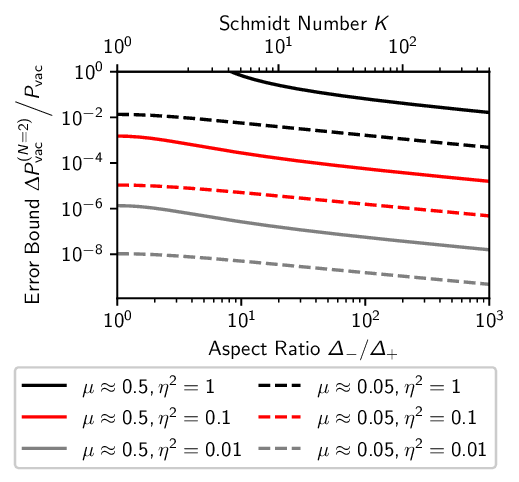}
    \caption{Relative error bound from \cref{eq:FredholmDeterminantExpansion_RelativeError} for the vacuum detection probabilities of a \typeII SPDC state with a Gaussian JSA when truncating the logarithm series in the determinant expansion after order ${N = 2}$. Different values of the total transmission $\eta^2$ and the mean number of generated photon pairs $\mu$ are considered.}
    \label{fig:RelativeError_Determinant}
\end{figure}

To quantify the quality and applicability of this approximation, in \cref{sec:Appendix_FredholmDeterminantApproximation} we derive an upper bound to \cref{eq:RelativeErrorVacuumProbability_Definition}, i.e.\ the relative error of the vacuum detection probability introduced by truncating the expansion in \cref{eq:FredholmDeterminant_LogarithmExpansion} after order $N$. \Cref{fig:RelativeError_Determinant} shows this bound for a \typeII SPDC state with a JSA given by a 2D~Gaussian as a function of the ratio of the standard deviations in anti-diagonal ($\varDelta_-$) and diagonal ($\varDelta_+$) direction for the lowest-order approximation $N=2$.\nolinebreak\footnote{For a 2D Gaussian JSA, the Schmidt decomposition in \cref{eq:SchmidtDecomposition} admits an analytic solution with the coefficients ${\lambda_j = (1-\zeta^2) \zeta^{2(j-1)}}$ and Schmidt number ${K = (1 + \zeta^2)/(1 - \zeta^2)}$, where ${\zeta = (\varDelta_-/\varDelta_+ - 1)/(\varDelta_-/\varDelta_+ + 1)}$~\cite{Mauerer_2009}.} Here, $\mu$ represents the mean number of generated photon pairs and $\eta$ is the maximum field transmittivity over all modes, i.e. the amplitude transmission factor that every mode experiences (cf. \cref{sec:Appendix_FredholmDeterminantApproximation}). The decrease of the relative error with an increasing aspect ratio and decreasing $\mu$ and $\eta$ can be observed, showing that even the lowest-order yields a sufficiently well approximation to the detection probabilities over a large regime of practically relevant parameters. Note that the emphasis here is to provide an upper bound to the relative error that can be computed efficiently even for states exhibiting strong spectral entanglement, however in most cases is not very tight. Thus, the actual error is expected to be notably smaller than the bound provided here.

\subsubsection{Compression of the Determinant}
A very common situation is that of some modes initially containing the vacuum state, e.g.\ due to beam splitters containing vacuum in one of their inputs. This can be used to effectively reduce the dimension of the covariance matrix and therefore significantly increase the computational performance when implementing the corresponding operations. Consider a system containing $M$ discrete modes, with ${M_0 = M - M' \geq 0}$ modes containing vacuum after all active transformations have been applied. Choosing the order of basis elements such that all vacuum modes are listed last, all transformations act according to
\begin{equation}\label{eq:ReducedTransformations}
    \bm{S} \mleft( \bm{\varGamma} \oplus \bm{0}_{2 M_0\times 2 M_0} \mright) \bm{S}^\dag = \bm{s} \bm{\varGamma} \bm{s}^\dag \,,
\end{equation}
where $\bm{\varGamma}$ is the ${2M' \times 2M'}$ renormalized covariance matrix describing only the modes initially not containing vacuum, $\bm{S}$ is the total ${2M \times 2M}$ transformation matrix, and $\bm{s}$ is a ${2M \times 2M'}$ reduced transformation matrix containing only the first $2M'$ columns of $\bm{S}$. Naively, the total transformation $\bm{S}$ is the composition of many subsequent $2M \times 2M$ transformations describing the components present in the setup. The total transformation $\bm{s}$ is obtained by multiplying ${2M \times 2M}$ matrices with ${2M \times 2M'}$ matrices, resulting in a significant computational advantage for large $M_0$.

Furthermore, this allows us to apply the analogue of Sylvester's determinant theorem\nolinebreak\footnote{Sylvester's determinant theorem states that for two matrices $\bm{A}_{m \times n}$ and $\bm{B}_{n \times m}$ it holds ${\det(\mathds 1_m + \bm A \bm B) = \det(\mathds 1_n + \bm B \bm A)}$~\cite{Pozrikidis_2014}. The continuous analogue to this statement used here can be inferred from the cyclic permutation under the trace in \cref{eq:FredholmDeterminant_LogarithmExpansion}.} for trace-class operators~\cite{Gohberg_2000} to extract the detection probabilities from the determinant of a smaller matrix:
\begin{equation}\label{eq:DeterminantSylvester}
    \det\mleft( \mathds{1}_{2M} + \bm{P} \bm{s} \bm{\varGamma} \bm{s}^\dag \bm{P} \mright) = \det\mleft( \mathds{1}_{2M'} + \bm{s}^\dag \bm{P} \bm{s} \bm{\varGamma} \mright) \,.
\end{equation}
The term ${\bm{s}^\dag \bm{P} \bm{s}}$ leads to the contributions of the different modes being added up before applying the determinant. For large matrices, the number of numerical operations to compute the determinant via the LU~decomposition scales approximately with the third power of the matrix dimension~\cite{Trefethen_1997}, thus the expected gain in performance amounts to a factor of $(M/M')^3$. For example, the simulation of the QKD system we present in \refcite{Kleinpass_2024_partII} requires ${M = 12}$ discrete modes, however, initially only ${M' = 2}$ discrete modes (one for each party) contain a non-zero intensity. By using \cref{eq:DeterminantSylvester}, the size of the matrix for which the determinant is to be computed reduces by a factor of~6, resulting in $(M/M')^3 = 216$.

The steps to compute the detection probabilities can be summarized as follows:
First, construct the renormalized covariance after application of the last active transformation. Then, identify all remaining passive transformations describing the setup under consideration to compute the total transformation $\bm{s}$ analytically on the block-matrix level for the discrete modes only. Compute the analytic expression of ${\bm{s}^\dag \bm{P} \bm{s} \bm{\varGamma}}$ depending on the modes over which a detection is performed, specified by $\bm{P}$.
These steps can be simplified by using software for symbolic computations when many transformations or many discrete DOFs are involved. Finally, translate the products between the different blocks to integral transforms to obtain the detection probabilities in terms of the continuous DOFs. We demonstrate the procedure for two entanglement-based QKD systems in the second part of this series, \refcite{Kleinpass_2024_partII}.

\section{Highly Entangled Biphoton States}\label{sec:HighlyEntangledStates}

In \cref{sec:Biphoton_States}, the representation of biphoton states in the continuous covariance formalism is discussed. The series expansion of the renormalized covariance as well as an easily computable error bound are presented in \cref{sec:Approximation_of_the_Covariance}, before explicitly examining the two lowest-order approximations in \cref{sec:Poisson_Approximations,sec:Hermite_Approximations}. 

\subsection{Covariance Representation of Biphoton States}\label{sec:Biphoton_States}

For simplicity, photon pair generations in wave guide structures supporting only one spatial mode are considered. In the undepleted pump approximation, the state of the generated photon pairs after tracing out the pump signal reads~\cite{Yang_2008, Mauerer_2009, Quesada_2014, Thomas_2021}
\begin{equation}\label{eq:SPDC_State}
    \ket{\psi} = \exp\mleft( \frac{C}{2} \! \int \!\! \dl\omega_\ups \dl\omega_\upi \psi(\omega_\ups, \omega_\upi) \hat{a}_\nu^\dag(\omega_\ups) \hat{a}_\rho^\dag(\omega_\upi) - \Hc \! \mright) \ket{0} \,,
\end{equation}
where $\nu$ and $\rho$ label the polarization modes of the signal and idler photons and the coefficient $C$ incorporates all constant factors such as the intensity of the pump field and the strength of the non-linear interaction. The only non-vanishing commutator is ${[\hat{a}_\nu(\omega_\ups), \hat{a}_\rho^\dag(\omega_\upi)] = \delta_{\nu\rho} \delta(\omega_\ups-\omega_\upi)}$. 

A process is called \typezero or \typeI (which we will summarize as \typezeroI) if both generated photons share the same polarization, i.e.\ ${\nu = \rho}$ in \cref{eq:SPDC_State}. Due to the waveguide supporting only one spatial mode, this leads to signal and idler being indistinguishable, i.e. featuring a symmetric JSA  ${\psi(\omega_\ups,\omega_\upi) = \psi(\omega_\upi,\omega_\ups)}$. If they are polarized orthogonally to each other, i.e.\ $\nu \neq \rho$, the process is called \typeII.

The joint spectral density (JSD) is the squared modulus of the JSA:
\begin{equation}
    \Psi(\omega_\ups, \omega_\upi) = \abs{\psi(\omega_\ups, \omega_\upi)}^2 \,.
\end{equation}
It describes the bivariate spectral probability density of the generated photons. As all constant factors are absorbed into $C$ in \cref{eq:SPDC_State}, the JSD is normalized to
\begin{equation}\label{eq:JSA_JSD_normalization}
    \int \dl\omega_\ups \dl\omega_\upi \Psi(\omega_\ups, \omega_\upi) = 1 \,.
\end{equation}
The signal and idler spectral densities are given by the marginal distributions of the JSD:
\begin{subequations}
    \begin{IEEEeqnarray}{rCl}
        \Psi_\ups(\omega_\ups) &=& \int \Psi(\omega_\ups, \omega_\upi) \dl \omega_\upi \,, \IEEElabel{eq:marginal_spectrum_A}\IEEEeqnarraynumspace\\
        \Psi_\upi(\omega_\upi) &=& \int \Psi(\omega_\ups, \omega_\upi) \dl \omega_\ups \,. \IEEElabel{eq:marginal_spectrum_B}
        \IEEEeqnarraynumspace
    \end{IEEEeqnarray}
\end{subequations}

A common approach to simplify the biphoton state from \cref{eq:SPDC_State} for sufficiently small pump powers is to truncate the series expansion after the linear term~\cite{Keller_1997, Grice_1997, Law_2000, Mikhailova_2008, Lee_2014, Liu_2020, Phehlukwayo_2020, Dorfman_2021}
\begin{equation}\label{eq:SPDC_linear}
    \ket{\psi} \propto \mleft( \ket{0} + \frac{C}{2} \int \dl\omega_\ups \dl\omega_\upi \psi(\omega_\ups,\omega_\upi) \hat{a}_\nu^\dag(\omega_\ups) \hat{a}_\rho^\dag(\omega_\upi) \ket{0} \mright)\,,
\end{equation}
neglecting the possibility of multiple photon pairs being created simultaneously. 

To include all orders of multi-pair events, the Gaussian state formalism may be employed. The quadratic unitary transformation in \cref{eq:SPDC_State} is re-written in the form of \cref{eq:SymplecticTransformationFromHamiltonian} to obtain
\begin{IEEEeqnarray}{c"c}\label{eq:Z_definition}
    \bm{Z}^{(\upI)} = C \mqty{ 0 & \bm{\psi} \\ \bm{\psi}^\dag & 0 } \,, &
    \bm{Z}^{(\upII)} = \frac{C}{2} 
    \mqty{
        0 & 0 & 0 & \bm{\psi}\\
        0 & 0 & \bm{\psi}^\upT & 0\\
        0 & \bm{\psi}^* & 0 & 0\\
        \bm{\psi}^\dag & 0 & 0 & 0
    } \,, \IEEEnonumber\\
\end{IEEEeqnarray}
for \typezeroI and \typeII processes, respectively, where $\bm{\psi}$ is the integral operator with kernel $\psi(\omega_\ups,\omega_\upi)$. The state is constructed by applying $\bm{Z}$ to the vacuum, such that, according to \cref{eq:CovarianceDisplacement_SymplecticTransformation}, the covariance is given by
\begin{equation}\label{eq:Covariance_SPDC_Exponential}
    \bm{\gamma} = \exp(2\bm{Z}) \,.
\end{equation}
The Schmidt decomposition~\cite{Law_2000, Lamata_2005, Mauerer_2009} of the JSA reads
\begin{equation}\label{eq:SchmidtDecomposition}
    \psi(\omega_\ups,\omega_\upi) = \sum_j \sqrt{\lambda_j} u_j(\omega_\ups) v_j^*(\omega_\upi)\,,
\end{equation}
where the Schmidt modes $\{u_j(\omega)\}_j$ and $\{v_j(\omega)\}_j$ are two sets of orthonormal functions and the Schmidt coefficients $\{\sqrt{\lambda_j}\}_j$ are real, non-negative and enumerated in decreasing order ${\sqrt{\lambda_j} \geq \sqrt{\lambda_{j+1}}}$. The normalization in \cref{eq:JSA_JSD_normalization} is represented by ${\sum_j \lambda_j = 1}$.

The magnitude of the entanglement can be quantified by the Schmidt number ${K = 1/\sum_j \lambda_j^2}$~\cite{Fedorov_2006, Mikhailova_2008, Mauerer_2009, Horoshko_2018}. Its minimum value ${K = 1}$ implies that ${\lambda_1 = 1}$ is the only non-zero Schmidt coefficient, corresponding to the JSD factorizing as ${\Psi(\omega_\ups,\omega_\upi) = \Psi_\ups(\omega_\ups) \Psi_\upi(\omega_\upi)}$, rendering the two photons independent from each other without any spectral entanglement between them. For a given number of non-zero Schmidt coefficients $J$, the Schmidt number takes its maximum value $K=J$ when all contributing coefficients are equal, i.e.\ ${\lambda_j = 1/J}$ for ${j \leq J}$. Such a state is called maximally entangled.

In analogy to the singular value decomposition of a matrix~\cite{Trefethen_1997}, the Schmidt decomposition of the Hilbert-Schmidt operator $\bm{\psi}$ with kernel $\psi(\omega_\ups,\omega_\upi)$ can be written compactly as ${\bm{\psi} = \bm{U} \bm{\Sigma} \bm{V}^\dag}$, with the diagonal matrix $\bm{\Sigma}$ containing the  singular values, that is, the Schmidt coefficients $\sqrt{\lambda_j}$. The columns of the unitary matrices $\bm{U}$ and $\bm{V}$ are the corresponding Schmidt modes of $\bm{\psi}$. This can be used to write the covariance in the form
\begin{subequations}\label{eq:Covariance_Type0TypeII}
    \begin{IEEEeqnarray}{rCl}
        \bm{\gamma}^{(\upI)} &=& 
        \mqty{
            \bm{U} \cosh (\bm{\sigma}^{(\upI)}) \bm{U}^\dag &
            \bm{U} \sinh(\bm{\sigma}^{(\upI)}) \bm{V}^\dag\\
            \bm{V} \sinh(\bm{\sigma}^{(\upI)}) \bm{U}^\dag &
            \bm{V} \cosh(\bm{\sigma}^{(\upI)}) \bm{V}^\dag
        } \,, \IEEElabel{eq:Covariance_Type0} \IEEEeqnarraynumspace\\[5pt]
        \bm{\gamma}^{(\upII)} &=&
        \mqty{
            \bm{U} \cosh(\bm{\sigma}^{(\upII)}) \bm{U}^\dag & \bm{U} \sinh(\bm{\sigma}^{(\upII)}) \bm{V}^\dag\\
            \bm{V} \sinh(\bm{\sigma}^{(\upII)}) \bm{U}^\dag & \bm{V} \cosh(\bm{\sigma}^{(\upII)}) \bm{V}^\dag
        } \bigoplus \text{c.c.} \,, \IEEEnonumber\IEEElabel{eq:Covariance_TypeII}\\
    \end{IEEEeqnarray}
\end{subequations}
for \typezeroI and \typeII processes, respectively, with the squeezing parameters
\begin{IEEEeqnarray}{c"c}\label{eq:SqueezingParameters_Definition}
    \bm{\sigma}^{(\upI)} = 2C \bm{\Sigma} \,,&
    \bm{\sigma}^{(\upII)} = C \bm{\Sigma} \,.
    \IEEEeqnarraynumspace
\end{IEEEeqnarray}
In the low-gain regime, where ${\cosh(\bm{\sigma}) \approx \mathds{1} + \bm{\sigma}^2/2}$, the corresponding mean numbers of generated photons pairs $\mu$ are given by
\begin{IEEEeqnarray}{c"c}\label{eq:MeanPhotonNumber_Poisson}
    \mu^{(\upI)} \approx C^2/2 \,, &
    \mu^{(\upII)} \approx C^2/4 \,.
    \IEEEeqnarraynumspace
\end{IEEEeqnarray}
For \typeII processes, the basis elements in \cref{eq:Covariance_TypeII} have been reordered according to
\begin{equation}\label{eq:BasisElementsReordering}
    \hat{\bm{\avec}} \to \big( \hat{a}(\omega), \hat{b}^\dag(\omega), \hat{a}^\dag(\omega), \hat{b}(\omega) \big)^\upT \,,
\end{equation}
where $\hat{a}(\omega)$ and $\hat{b}(\omega)$ correspond to the orthogonal polarization modes of signal and idler, respectively.\nolinebreak\footnote{The shape  of $\bm{\gamma}^{(\upII)}$ is a consequence of the generation of signal and idler photons in orthogonal polarization modes. When separating both photons of a \typezeroI process by using wavelength-division demultiplexing, the covariance takes on the same shape if the frequency channels are chosen such that both photons can never end up in the same channel (see \refcite{Kleinpass_2024_partII}).} \Cref{sec:Appendix_Eigenvalues_and_Generating_Functions} shows how the eigenvalues of the renormalized covariance as well as the generating functions of the processes are obtained from the squeezing parameters.

\subsection{Approximation of the Renormalized Covariance}\label{sec:Approximation_of_the_Covariance}

For a large amount of spectral entanglement, the Schmidt decomposition of the JSA in \cref{eq:Covariance_Type0TypeII} becomes very challenging to evaluate in practice. Instead, the renormalized covariance can be approximated by truncating the series expansion of the exponential in \cref{eq:Covariance_SPDC_Exponential} at some sufficiently large order $N$:
\begin{equation}\label{eq:RenormalizedCovariance_SeriesExpansion}
    \bm{\varGamma}_N \approx \sum_{n=1}^N \frac{(2\bm{Z})^n}{2 n!} \,.
\end{equation}
In \cref{sec:Error_Bound_Renormalized_Covariance} we show that the relative truncation error w.r.t.\ the trace norm\footnote{The trace norm of an operator $A$ is given by $\norm{A}_{\mathrm{Tr}} = \Tr(\sqrt{A^\dagger A})$.} is bounded by
\begin{equation}\label{eq:RenormalizedCovarianceApproximation_RelativeError}
    \frac{\norm{\bm{\varGamma} - \bm{\varGamma}_N}_{\mathrm{Tr}}}{\norm{\bm{\varGamma}}_{\mathrm{Tr}}} \leq
    \mleft\{ \begin{aligned}
        &\frac{\sum_{j=1}^M \mleft[ \sinh(\sigma_j) - \mathfrak{s}_N(\sigma_j) \mright]}{\sum_{j=1}^M \sinh(\sigma_j)} & &\text{for $N$ even,}\\
        &\frac{\sum_{j=1}^M \mleft[ \cosh(\sigma_j) - \mathfrak{c}_N(\sigma_j) \mright]}{\sum_{j=1}^M \sinh(\sigma_j)} & &\text{for $N$ odd,}
    \end{aligned} \mright.
\end{equation}
for all $M \leq J$, where
\begin{IEEEeqnarray}{c"c}
    \mathfrak{c}_N(x) = \sum_{\substack{n\\\mathclap{\{ 0 \leq 2n \leq N \}}}} \frac{x^{2n}}{(2n)!} \,, &
    \mathfrak{s}_N(x) = \sum_{\substack{n\\\mathclap{\{ 0 \leq 2n+1 \leq N \}}}} \frac{x^{2n+1}}{(2n+1)!} \,.
    \IEEEeqnarraynumspace
    \IEEElabel{eq:truncated_sinh_cosh_definiton}
\end{IEEEeqnarray}
are the contributions to the $\cosh(x)$ and $\sinh(x)$ series when truncating the expansion of $\exp(x)$ at order $N$, respectively. This means, that the relative error is bounded by the $M$ largest squeezing parameters, where equality holds for $M = J$.

The relative error in \cref{eq:RenormalizedCovarianceApproximation_RelativeError} takes its maximum value in the non-entangled case, i.e. when all of the state's energy is concentrated within one Schmidt mode. For an increasing amount of entanglement, more and more Schmidt modes become relevant and it becomes more and more expensive to perform an explicit Schmidt decomposition. At the same time, the relative error in \cref{eq:RenormalizedCovarianceApproximation_RelativeError} decreases the more uniformly the energy is distributed amongst an increasing number of contributing Schmidt modes. Therefore, the more expensive it becomes to perform an explicit Schmidt decomposition, the faster the sum in \cref{eq:RenormalizedCovarianceApproximation_RelativeError} converges and the easier it becomes to use this approximation.

\begin{figure}
    \centering
    \includegraphics{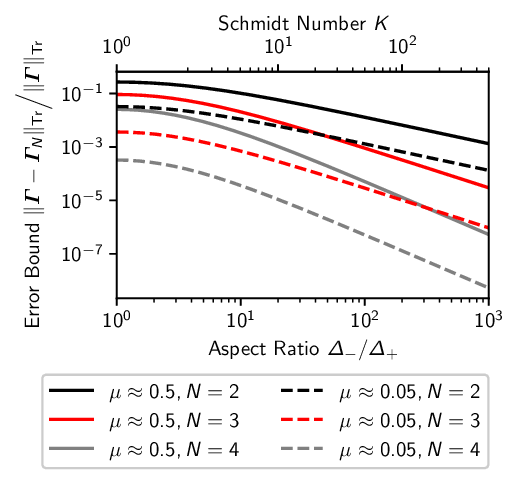}
    \caption{Relative error bound from \cref{eq:RenormalizedCovarianceApproximation_RelativeError} for the $N$-th order approximation of the renormalized covariance of a \typeII SPDC state with Gaussian JSA w.r.t. the trace norm. Different expansion orders $N$ and mean numbers of generated photon pairs $\mu$ are considered.}
    \label{fig:RelativeError_Covariance}
\end{figure}

In practice, to obtain a bound for the relative approximation error without performing a full Schmidt decomposition, it is sufficient to compute the $M$ largest squeezing parameters, which can be done by discretizing $\psi(\omega_\ups, \omega_\upi)$ and computing a truncated singular value decomposition of the resulting matrix~\cite{Golub_1965,Hochstenbach_2001,Halko_2011,Wu_2015}. \Cref{fig:RelativeError_Covariance} shows the upper bound of the relative error for $M=1$, i.e. only using the largest squeezing parameter $\sigma_1$ and for different values of $\mu$ and $N$ over the aspect ratio ${\varDelta_-/\varDelta_+}$ of a 2D~Gaussian JSA~\cite{Mauerer_2009}.

\subsection{Bivariate Poisson Approximation}\label{sec:Poisson_Approximations}

The lowest-order expansions of the covariance and determinant series provides further insights into the physical meaning of the different expansion orders. For the covariance, the approximation orders $N=0$ and $N=1$ in \cref{eq:RenormalizedCovariance_SeriesExpansion} yield no contribution to the expected number of photons, such that the lowest non-trivial expansion order is given by $N=2$:
\begin{equation}\label{eq:second_order_expansion_of_renormalized_covariance}
    \bm{\varGamma} \approx \bm{Z} + \bm{Z}^2 \,.
\end{equation}
Similarly, as discussed in \cref{sec:Evaluating_the_Probabilities}, at least a second order approximation of the logarithm is required to account for correlations. In the presence of any block-diagonal transformations, which we represent as frequency-dependent losses $\bm{\eta}$, i.e. ${\bm{\varGamma} \to \bm{\eta} \bm{\varGamma} \bm{\eta}}$, the resulting generating function reads
\begin{equation}\label{eq:Fredholm_determinant_second_order_approximation}
    G^{(N=2)}(\bm{w}) \approx \exp\mleft( - \frac{\Tr(\bm{W} \bm{\eta} \bm{\varGamma} \bm{\eta})}{2} + \frac{\norm{\bm{W} \bm{\eta} \bm{\varGamma} \bm{\eta}}_{\mathrm{HS}}^2}{4} \mright) \,.
\end{equation}
Due to the anti-diagonal structure of $\bm{Z}$, only even orders of $\bm{Z}$ can contribute. Thus, the second term in \cref{eq:Fredholm_determinant_second_order_approximation} is composed of two contributions after inserting \cref{eq:second_order_expansion_of_renormalized_covariance}:
\begin{equation}\label{eq:Second_Order_Trace}
    \norm{\bm{W}_{\bm{\eta}} \bm{\varGamma}}_{\mathrm{HS}}^2 = \underbrace{\norm{\bm{W}_{\bm{\eta}} \bm{Z}}_{\mathrm{HS}}^2}_{\mathcal{O}(\psi^2)} + \underbrace{\norm{\bm{W}_{\bm{\eta}} \bm{Z}^2 }_{\mathrm{HS}}^2}_{\mathcal{O}(\psi^4)} \,.
\end{equation}
For a large amount of spectral entanglement or a sufficiently small mean number of photon pairs, the $\mathcal{O}(\psi^4)$ term can be neglected, such that
\begin{equation}\label{eq:PoissonApproximation_PGF_Z}
    G^{(\mathrm{Poisson})}(\bm{w}) = \exp\mleft[ -\frac{1}{4} \Tr\mleft( \bm{Z}^2 - \mleft[ \mleft( \Id - \bm{W}_{\bm{\eta}} \mright) \bm{Z} \mright]^2 \mright) \mright] \,,
\end{equation}
where $\bm{W}_{\bm{\eta}} = \bm{\eta} \bm{W} \bm{\eta}$. This can be rewritten according to
\begin{equation}\label{eq:PoissonApproximation_PGF}
    G^{(\mathrm{Poisson})}(\bm{w}) = \upe^{-\mu \mleft( w_\ups p_\ups(I_\ups) + w_\upi p_\upi(I_\upi) - w_\ups w_\upi p_{\ups,\upi}(I_\ups \cap I_\upi) \mright)} \,,
\end{equation}
where $\mu$ is given by \cref{eq:MeanPhotonNumber_Poisson}.\\
For \typeII processes,
\begin{subequations}\IEEEyesnumber \label{eq:Probability_SignalIdlerDetection}
    \begin{IEEEeqnarray}{rCl}
        p_\ups(I_\ups) &=& \Tr\mleft( \bm{W}_{\bm{\eta}} \bm{\psi} \bm{\psi}^\dag \mright) = \int_{I_\ups} \dl \omega_\ups \, \eta_\ups^2(\omega_\ups) \Psi_\ups(\omega_\ups) \,, \IEEEeqnarraynumspace \\
        p_\upi(I_\upi) &=& \Tr\mleft( \bm{W}_{\bm{\eta}} \bm{\psi}^\dag \bm{\psi} \mright) = \int_{I_\upi} \dl \omega_\upi \, \eta_\upi^2(\omega_\upi) \Psi_\upi(\omega_\upi) \,, \IEEEeqnarraynumspace
    \end{IEEEeqnarray}
\end{subequations}
and
\begin{equation}\label{eq:Probability_CoincidenceDetection}
    p_{\ups,\upi}(I_\ups \cap I_\upi) = \Tr\mleft( \bm{W}_{\bm{\eta}} \bm{\psi} \bm{W}_{\bm{\eta}} \bm{\psi}^\dag \mright) = \int_{I_\ups} \dl\omega_\ups \int_{I_\upi} \dl\omega_\upi \, \eta_\ups^2(\omega_\ups) \eta_\upi^2(\omega_\upi) \Psi(\omega_\ups,\omega_\upi)
\end{equation}
are the probabilities of detecting signal or idler in their corresponding interval and coincidentally detecting both photons in their intervals, respectively, conditioned on the generation of exactly one photon pair.\\
For \typezeroI SPDC, due to the indistinguishability of signal and idler, we have $I_\ups = I_\upi = I$ and therefore
\begin{equation}
    p_\ups(I_\ups) = p_\upi(I_\upi) = \Tr\mleft( \bm{W}_{\bm{\eta}} \bm{\psi} \bm{\psi}^\dag \mright) = \int_I \dl\omega \eta^2(\omega) \Psi(\omega) \,,
\end{equation}
as well as
\begin{equation}
    p_{\ups,\upi}(I_\ups \cap I_\upi) = \Tr\mleft( \bm{W}_{\bm{\eta}} \bm{\psi} \bm{W}_{\bm{\eta}} \bm{\psi}^\dag \mright) = \int_I \dl\omega \int_I \dl\omega' \eta^2(\omega) \eta^2(\omega') \Psi(\omega, \omega') \,.
\end{equation}
In many applications featuring \typezeroI processes, signal and idler would be distinguished after separating both generated photons by their frequency, e.g. by introducing a wavelength division demultiplexer with two non-overlapping channels for signal and idler: ${\eta^2(\omega) \to \eta_\ups^2(\omega) + \eta_\upi^2(\omega)}$. However, depending on the choice of the frequency channels and the shape of the JSA, there will be a possibility of both photons of a pair ending up in the same channel. This will be discussed in the second part of this series, \refcite{Kleinpass_2024_partII}. If the channels are chosen such that both photons will never end up in the same channel, signal and idler are well-defined and the resulting expressions are equal to the ones for \typeII processes in \cref{eq:Probability_SignalIdlerDetection} and \cref{eq:Probability_CoincidenceDetection}.

The covariance may always be transformed into the time domain via the symplectic Fourier transform in \cref{eq:Symplectic_FourierTransform}, permitting the consideration of detections in time instead of frequency intervals. Using \cref{eq:Interlacing}, the additional relative error of neglecting the $\mathcal{O}(\psi^4)$ term in \cref{eq:Second_Order_Trace} is bounded by
\begin{equation}
    \frac{\Updelta P_{\vac}^{(\mathrm{Poisson})}}{P_{\vac}^{(N=2)}} \leq 
    \mleft\{\begin{aligned}
        &1 - \exp\mleft(-\frac{\eta^4}{2 K} C^4 \mright) & &\text{for \typezeroI,}\\
        &1 - \exp\mleft(-\frac{\eta_\ups^4 + \eta_\upi^4}{32 K} C^4 \mright) & &\text{for \typeII,}
    \end{aligned}\mright.
\end{equation}
where ${\Updelta P_{\vac}^{(\mathrm{Poisson})} = \abs{P_{\vac}^{(N=2)} - P_{\vac}^{(\mathrm{Poisson})}}}$ and $\eta^2, \eta_\ups^2, \eta_\upi^2$ are the maxima of the corresponding intensity transmission functions.

Evidently, \cref{eq:PoissonApproximation_PGF} is the generating function of a bivariate Poisson distribution~\cite{Campbell_1934, Kawamura_1973}. Therefore, the signal and idler marginals as well as the pair distribution are Poissonian with means ${\mu p_\ups(I_\ups)}$, ${\mu p_\upi(I_\upi)}$ and $\mu p_{\ups,\upi}(I_\ups \cap I_\upi)$, respectively. The vacuum probabilities in this approximation are given by
\begin{equation}\label{eq:VacuumProb_Poisson}
    P_{\vac}^{(\mathrm{Poisson})} = G^{(\mathrm{Poisson})}(1, 1) =  \upe^{-\mu p_{\ups,\upi}(I_\ups \cup I_\upi)} \,,
\end{equation}
where ${p_{\ups, \upi}(I_\ups \cup I_\upi) = p_\ups(I_\ups) + p_\upi(I_\upi) - p_{\ups,\upi}(I_\ups \cap I_\upi)}$ is the probability of finding at least one of the photons in the corresponding intervals.

This leads to a very intuitive interpretation of this approximation: As discussed in \cref{sec:Biphoton_States}, the photon-pair statistics are expected to be Poissonian in the limit of an infinite amount of entanglement. Thus, combining the series expansions in \cref{eq:RenormalizedCovariance_SeriesExpansion} and \cref{eq:FredholmDeterminant_LogarithmExpansion} yields an approximation of the state in terms of an infinitely entangled state with Poissonian pair statistics and higher-order corrections.

\begin{figure}
    \centering
    \includegraphics{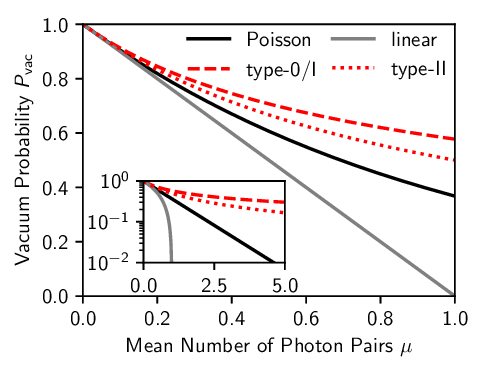}
    \caption{Vacuum detection probability for the Poisson approximation, the non-entangled case with only one contributing Schmidt mode for \typezeroI and \typeII processes and the linear approximation from \cref{eq:SPDC_linear}.}
    \label{fig:PhotonPairStatistics}
\end{figure}

Note that the bivariate Poisson approximation is always more accurate than the one-pair approximation in \cref{eq:SPDC_linear}, independent of the shape of the JSA and the mean number of photon pairs, even though it requires the evaluation of the same integrals. In the one-pair approximation, the coincidence vacuum probability is given by ${P_{\vac}^{(\mathrm{linear})} = 1 - \mu p_{\ups,\upi}(I_\ups \cup I_\upi)}$, which is the first-order expansion of \cref{eq:VacuumProb_Poisson}. Thus, as depicted in \cref{fig:PhotonPairStatistics}, it is always smaller than the lower bound of the vacuum detection probabilities given in \cref{eq:VaccuumProbability_Bounds}.

\subsection{Bivariate Hermite Approximation}\label{sec:Hermite_Approximations}

The bivariate Poisson approximation is valid if different pairs are generated independently from each other such that the multi-pair statistics can be inferred from the state of a single-pair. Therefore, it can be regarded as the analogue of the linear approximation of \cref{eq:SPDC_State}, both requiring the evaluation of $\mathcal{O}(\psi^2)$ terms. Similarly, the quadratic approximation of \cref{eq:SPDC_State} describes the generation of two pairs through $\mathcal{O}(\psi^4)$ terms, including correlations between both pairs. Extending our series expansions correspondingly to $\mathcal{O}(\psi^4)$ terms refines the statistics, introducing pairwise correlations between all generated pairs. To include those terms, series expansions of the renormalized covariance
\begin{equation}
    \bm{\varGamma} \approx \bm{Z} + \bm{Z}^2 + \frac{2 \bm{Z}^3}{3} + \frac{\bm{Z}^4}{3}
\end{equation}
as well as the logarithm up to the fourth order need to be employed. The resulting generating function can be written as
\begin{equation}
    G^{(\mathrm{Hermite})}(\bm{w}) = \exp\bigg\{ -\frac{1}{8} \Tr\mleft( \bm{Z}^4 - \mleft[ \mleft( \Id - \bm{W}_{\bm{\eta}} \mright) \bm{Z} \mright]^4 \mright) - \frac{1}{4} \Tr\mleft[\mleft( \Id - \frac{2}{3} \bm{Z}^2 \mright) \mleft( \bm{Z}^2 - \mleft[ \mleft( \Id - \bm{W}_{\bm{\eta}} \mright) \bm{Z} \mright]^2\mright) \mright] \bigg\} \,.
\end{equation}
For frequency-independent losses, this can be simplified to\footnote{This corresponds to the $H_8(u,v)$ bivariate Hermite distribution introduced in \refcite{Kemp_1982},
\begin{equation}\label{eq:H8_GeneratingFunction}
        H_8(u, v) = \exp\big( b_1 w_s + b_2 w_s^2 + b_3 w_i + b_4 w_i^2  + b_5 w_s w_i + b_6 w_s^2 w_i + b_7 w_s w_i^2 + b_8 w_s^2 w_i^2 \big) \,, 
\end{equation}
with $w_\ups = u-1$, $w_\upi = v-1$ and parameters ${b_1 = -\eta_\ups^2 \mu}$, ${b_2 = \eta_\ups^4 \varepsilon^2/2}$, ${b_3 = -\eta_\upi^2 \mu}$, ${b_4 = \eta_\upi^4 \varepsilon^2/2}$, ${b_5 = \eta_\ups^2 \eta_\upi^2 (\mu + \varepsilon^2)}$, ${b_6 = -\eta_\ups^4 \eta_\upi^2 \varepsilon^2}$, ${b_7 = -\eta_\ups^2 \eta_\upi^4 \varepsilon^2}$ and ${b_8 = \eta_\ups^4 \eta_\upi^4 \varepsilon^2/2}$.}
\begin{equation}
    G^{(\mathrm{Hermite})}(\bm{w}) = \exp\bigg[ -\frac{\varepsilon^2}{2} \mleft( 1 - \mleft( 1 - \eta_\ups^2 w_\ups \mright)^2 \mleft(1 - \eta_\upi^2 w_\upi \mright)^2 \mright) - \mleft( \mu - \varepsilon^2 \mright) \mleft( 1 - \mleft( 1 - \eta_\ups^2 w_\ups \mright) \mleft( 1 - \eta_\upi^2 w_\upi \mright) \mright) \bigg] \,.
\end{equation}
Again, for \typezeroI processes with indistinguishable signal and idler photons, we have $\eta_\ups = \eta_\upi = \eta$ and $w_\ups = w_\upi = w$.

Due to the increased amount of correlations, the mean number of photon pairs is transformed according to $\mu \to \mu + \epsilon^2/3$ w.r.t. the Poisson approximation in \cref{eq:MeanPhotonNumber_Poisson}, i.e.
\begin{IEEEeqnarray}{c"c}\label{eq:MeanPhotonNumber_Hermite}
    \mu^{(\upI)} = \frac{C^2}{2} + \frac{C^4}{6K} \,, &
    \mu^{(\upII)} = \frac{C^2}{4} + \frac{C^4}{48K} \,,
\end{IEEEeqnarray}
where
\begin{IEEEeqnarray}{c"c}
    \mleft( \varepsilon^{(\upI)} \mright)^2 = \frac{C^4}{2 K} \,, &
    \mleft( \varepsilon^{(\upII)} \mright)^2 = \frac{C^4}{16 K} \,,
\end{IEEEeqnarray}
is a measure of the strength of the two-pair correlations.\\
In this approximation, the pair generation is described as the sum of two independent Poisson processes, one generating photon pairs with mean $\mu - \varepsilon^2$, while the other one generates pairs of pairs with mean $\varepsilon^2/2$. An easy way of seeing this is to observe the marginal statistics of signal photon generations, i.e. setting $\eta_\ups = 1$ and $w_\upi = 0$, resulting in
\begin{equation}
    G^{(\mathrm{Hermite})}(w_\ups) = \upe^{ -(\mu - \varepsilon^2) w_\ups - \varepsilon^2 \mleft( 2 w_\ups - w_\ups^2 \mright)/2 } \,.
\end{equation}
This corresponds to the probability-generating function of a Hermite distribution~\cite{Kemp_1965, Kemp_1966}. The probability of $n$ pairs being generated is given by the $n$-th Hermite polynomial $H_n$ according to \cite{Kemp_1965}
\begin{equation}
    P(n) = \frac{\varepsilon^n}{\upi^n \sqrt{2^n} n!} \upe^{-(\mu - \varepsilon^2/2)} H_n\mleft( \frac{\upi (\mu-\varepsilon^2)}{\sqrt{2} \varepsilon} \mright) \,.
\end{equation}
Note that this only represents a valid probability distribution for $\mu \geq \varepsilon^2$. The correlations between different pairs can be quantified by the second-order correlation function \cite{Christ_2011}
\begin{equation}
    g^{(2)} = \frac{\langle \hat{n}^2 \rangle - \langle \hat{n} \rangle}{\langle \hat{n} \rangle^2} = 1 + \frac{\varepsilon^2}{\mu^2} \geq 1 \,,
\end{equation} with $\langle\hat n\rangle$ and $\langle \hat n^2 \rangle$ calculated from the moment-generating function ${M^{(\mathrm{Hermite})}(w) = G^{(\mathrm{Hermite})}(1-\upe^w)}$~\cite{Fitzke_2023}.

\begin{figure}
    \centering
    \includegraphics{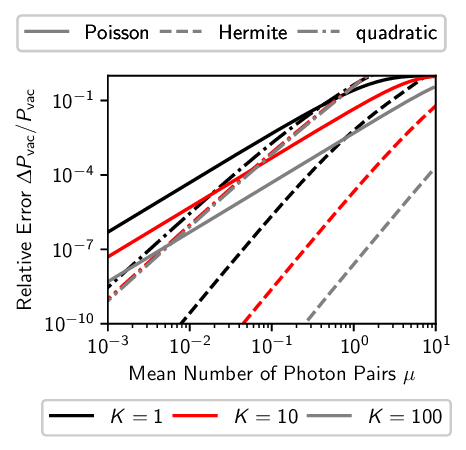}
    \caption{Relative error of the vacuum detection probability for \typeII SPDC employing the bivariate Poisson approximation, the bivariate Hermite approximation and the quadratic expansion of \cref{eq:SPDC_State} w.r.t. a process with Gaussian JSA.}
    \label{fig:Error_to_Gauss}
\end{figure}

In \cref{fig:Error_to_Gauss} the relative error of the vacuum probability for \typeII SPDC is presented for a process with a Gaussian JSA, comparing the bivariate Poisson approximation, the bivariate Hermite approximation and the quadratic two-pair expansion of \cref{eq:SPDC_State}. It has already been established that the linear expansion always performs worse than the bivariate Poisson approximation, thus it is omitted here. Although significantly easier to compute, the bivariate Poisson approximation outperforms the quadratic expansion in a large parameter regime, while the bivariate Hermite approximation outperforms the quadratic expansion for all examined parameters.

\FloatBarrier
\section{Summary and Conclusions}
Highly entangled biphoton states exhibit a large amount of contributing Schmidt modes in the decomposition of their JSA. This is usually accompanied by a large aspect ratio between the sum and difference frequencies of signal and idler photons. Both of these properties render the usual approach of computing the Schmidt decomposition by discretization and subsequent singular value decomposition computationally expensive. A commonly used alternative is to truncate the expansion of the squeezing operator after the term of second or fourth order in the JSA, as the inclusion of higher orders becomes increasingly more complex. This corresponds to single or two-pair generations but neglects all higher numbers of multi-pair events. These, however, can constitute a limiting factor to the performance of many systems and should therefore be properly accounted for.

In this publication, we discussed an approach that avoids the expensive computation of higher orders of the JSA whilst still accounting for higher orders of photon-pair generations. This is done by introducing series expansions of the renormalized covariance and the detection probabilities arising from the formalism of Gaussian states. An intuitive interpretation of our expansions is given in terms of correlations, where terms up to second order correspond not to a single, but a multitude of independent pair generations, while terms up to fourth order introduce correlations (bunching) between pairs of pairs. We showed that with equal numerical complexity, these expressions yield more accurate results for the detection probability than the usual one- and two-pair expansions for a wide range of mean photon numbers.

We discussed the covariance formalism of Gaussian states in the limit of a continuum of time or frequency modes and identified the renormalized covariance as the essential quantity of interest. It is a matrix whose rows and columns represent the discrete degrees of freedom~(DOFs) of the state while each of its elements is an integral operator with a kernel representing the distribution over the continuous DOFs. As the renormalized covariance is a trace class operator, the determinants arising in the expressions for the detection probabilities are well-defined. The separate treatment of discrete and continuous DOFs allows for a generalization of Gaussian transformations to a combination of matrix multiplications over the discrete DOFs and integrals over the continuous DOFs, as well as simple modeling of photon-number resolved detection.

Future work could apply our considerations to similar states with a non-zero displacement vector and extend the results to multi-dimensional continuous DOFs such as position and momentum, which is relevant for example to model the generation of photon pairs that are not confined to waveguide structures.

\appendix
\FloatBarrier

\section{Continuous-Mode Gaussian Transformations}\label{sec:Appendix_Gaussian_Transformations}

Gaussian transformations are transformations mapping one Gaussian state to another one. The respective symplectic transformations given by \cref{eq:SymplecticTransformationFromHamiltonian} need to be adapted to the continuous-mode setting by replacing the normal-mode operators in \cref{eq:NormalModeVector_Definition} with their continuous analogues and obtaining the corresponding transformations via \cref{eq:QuadraticUnitaryTransformation_Definition}. To simplify the notation, we make use of the invariance of \cref{eq:DetectionProbability_RenormalizedCovariance} under permutations of the basis elements and reorder them such that the modes affected by the transformation are listed first.

As described above, the elements of the transformation matrices are themselves linear integral operators. Most operators $\bm{f}$ of interest here are diagonal in the continuous-mode variables in the sense that their kernel is given by ${f(\omega,\omega') = f(\omega) \delta(\omega-\omega')}$. Such operators act as multiplication operators ${(\bm{f} \bm{x})(\omega) = f(\omega) x(\omega)}$ and for ease of notation we will denote them by simply writing $f(\omega)$ when they occur in block structures such as in \cref{eq:phase_shift_operator}.\\

\noindent \textbf{Phase Shift}

The transformation shifting the phase of all modes within one discrete DOF by $\phi(\omega)$ is given by
\begin{equation}\label{eq:phase_shift_operator}
    \bm{\PS} = \mqty{ \upe^{\upi \phi(\omega)} & 0\\ 0 & \upe^{-\upi \phi(\omega)} } \bigoplus \mathds{1} \,,
\end{equation}
meaning that the corresponding modes are simply multiplied with a frequency-dependent phase factor. To describe the phase acquired due to the propagation within some medium of length $L$, a Taylor expansion of the wave number $k(\omega)$ around the carrier frequency $\omega_0$ is performed:
\begin{equation}
    k(\omega) = \frac{n(\omega) \omega}{c} = \frac{n_0 \omega_0}{c} + \frac{n_{\mathrm{g}}}{c} \bar{\omega} + \frac{\beta}{2} \bar{\omega}^2+ \mathcal{O}(\bar{\omega}^3)\,,
\end{equation}
where ${\bar{\omega} = \omega-\omega_0}$. Neglecting $\mathcal{O}(\bar{\omega}^3)$ terms yields the phase
\begin{equation}
    \phi(\omega) = k(\omega) L = \phi_0 + \tau \bar{\omega} + \frac{\beta}{2} L \bar{\omega}^2 \,.
\end{equation}
The first term $\phi_0$ describes a constant phase shift ${\phi_0 = \omega_0 \tau n_0/n_{\mathrm{g}}}$, where $n_0$ and $n_{\mathrm{g}}$ are the refractive phase and group indices, respectively. The second term represents the time ${\tau= n_{\mathrm{g}} L / c}$ the signal takes to travel the distance $L$, and the third term describes a chirp, inducing chromatic dispersion according to the group velocity dispersion $\beta$.\\

\noindent \textbf{Fourier Transform}

The covariance can be transformed between the temporal and spectral domain by means of the symplectic Fourier transformation
\begin{equation}\label{eq:Symplectic_FourierTransform}
    \bm{F} = \mqty{ \bm{\mathcal{F}} & 0\\ 0 & \bm{\mathcal{F}}^* } \bigoplus \mathds{1} \,,
\end{equation}
where $\bm{\mathcal{F}}$ is the unitary Fourier transform, i.e.\ the integral operator with kernel
${\mathcal{F}(t, \omega) = \upe^{-\upi \omega t}/\sqrt{2\uppi}}$.\\

\noindent \textbf{Beam Splitter}

One of the most important transformations is the mixing of modes at a beam splitter, which can also be used to describe many important experimental imperfections such as losses and mode mismatches. In general, a frequency-dependent mode mixing between the modes of two discrete DOFs is described by the transformation matrix
\begin{equation}
    \bm{\BS} = \mqty{ \mathcal{T}(\omega) & \mathcal{R}(\omega)\\ -\mathcal{R}(\omega) & \mathcal{T}(\omega) }^{\oplus2} \bigoplus \mathds{1} \,,
\end{equation}
where $\mathcal{T}$ and $\mathcal{R}$ are the transmission and reflection coefficients with ${\mathcal{T}(\omega)^2 + \mathcal{R}(\omega)^2 = 1}$.\\

\noindent \textbf{Orthogonal Projections}

The detection statistics are typically considered only over a subset of modes. The modes of interest are selected by application of the orthogonal projection
\begin{equation}\label{eq:ProjectionContinuum}
    \bm{P} = \bigoplus_{m=1}^{M} \mleft[\rect_{I_m}(\omega)\mright]^{\oplus 2} \,,
\end{equation}
where $I_m$ is the interval over which the corresponding mode is observed and $\rect_{I_m}$ is a rectangular function spanning this interval either in the time or frequency domain. For ${I_m \to (-\infty, \infty)}$, the projection becomes the unit operator for the discrete DOF $m$ and, similarly, for ${I_m \to \emptyset}$ it becomes the zero operator.\\

\noindent \textbf{Loss Transformations}

Losses introduced by optical devices lower the system's total energy and are, therefore, no passive transformations. However, the loss transformation can be described by coupling the system to a number of auxiliary loss modes via virtual beam splitters~\cite{Takeoka_2015}. Those modes are never detected but always discarded by the projection in \cref{eq:ProjectionContinuum}. As the loss modes do not interact with the system anymore, the projection can be performed right away, simplifying the loss transformation to ${\bm{\varGamma} \to \bm{\eta} \bm{\varGamma} \bm{\eta}}$, where
\begin{equation}\label{eq:LossTransformation}
    \bm{\eta} = \mqty{ \eta(\omega) & 0\\ 0 & \eta(\omega) } \bigoplus \mathds{1} \,,
\end{equation}
with the field transmittivity factor ${\eta(\omega) \leq 1}$~\cite{Takeoka_2015}.

\section{Eigenvalues and Generating Functions}\label{sec:Appendix_Eigenvalues_and_Generating_Functions}
The eigenvalues $\{ \Lambda_{\pm j} \}_j$ of the renormalized covariance can be inferred from the squeezing parameters via
\begin{equation}\label{eq:CovarianceSPDC_Eigenvalues}
    \Lambda_{\pm j} = \frac{\upe^{\pm \sigma_j} - 1}{2} \,,
\end{equation}
ordered according to ${\Lambda_{-(j+1)} \leq \Lambda_{-j} \leq 0 \leq \Lambda_j \leq \Lambda_{j+1}}$. For \typeII processes, each eigenvalue is at least twofold degenerate. From the eigenvalues, an analytical expression for the generating function can be obtained. It reads
\begin{subequations}
    \begin{IEEEeqnarray}{rCl}
        G^{(\upI)}(w) &=& \prod_{j=1}^J \frac{1}{\sqrt{\cosh^2\frac{\sigma_j}{2} - (1-w)^2 \sinh^2\frac{\sigma_j}{2}}} \,, \IEEEeqnarraynumspace \\
        G^{(\upII)}(w_\ups, w_\upi) &=& \prod_{j=1}^J \prod_{\rho=\ups,\upi} \frac{1}{\sqrt{\cosh^2\frac{\sigma_j}{2} - (1-w_\rho)^2 \sinh^2\frac{\sigma_j}{2}}} \,, \IEEEnonumber\\
    \end{IEEEeqnarray}
\end{subequations}
where $J$ is the number of non-zero Schmidt coefficients of the JSA.

The vacuum probabilities are obtained by setting ${w = w_\ups = w_\upi = 0}$. For a given mean number of photon pairs $\mu$, it takes its maximum value for non-entangled states, where ${\sigma_j = \sigma_1 \delta_{1j}}$ and, by Jensen's inequality,\nolinebreak\footnote{Jensen's inequality~\cite{Jensen_1906} states that for a convex function $f(x)$ and ${c_j > 0}$: ${f(\sum_j c_j x_j/\sum_j c_j) \leq \sum_j c_j f(x_j)/\sum_j c_j}$.} it takes its minimum value for maximally entangled states, where ${\sigma_j = \sigma_{j+1}}$ for all ${j \leq J-1}$. Thus, the maximum and minimum vacuum detection probabilites are attained for one and infinitely many equally contributing Schmidt-modes, respectively:
\begin{IEEEeqnarray}{c"c}\label{eq:VaccuumProbability_Bounds}
    \frac{1}{\sqrt{1 + 2 \mu}} \geq P_{\vac}^{(\upI)} \geq \upe^{-\mu} \,, &
    \frac{1}{1 + \mu} \geq P_{\vac}^{(\upII)} \geq \upe^{-\mu} \,,
    \IEEEeqnarraynumspace
\end{IEEEeqnarray}
where ${\lim_{J \to \infty} \cosh^{J}(x/\sqrt J) = \exp(x^2/2)}$ and \cref{eq:MeanPhotonNumber_Poisson} were used. This also reflects the fact that in the limit of an infinite amount of contributing Schmidt modes ${J \to \infty}$, a maximally entangled state approaches Poissonian pair statistics~\cite{Mauerer_2009}.

\section{Error Bound for the Renormalized Covariance Series Approximation}\label{sec:Error_Bound_Renormalized_Covariance}
The relative error ${\norm{\bm{\varGamma} - \bm{\varGamma}_N}_{\mathrm{Tr}}/\norm{\bm{\varGamma}}_{\mathrm{Tr}}}$ introduced by truncating the series expansion of the renormalized covariance in \cref{eq:RenormalizedCovariance_SeriesExpansion} is given by
\begin{equation}
    \frac{\sum_j \abs*{\upe^{\sigma_j} - \sum_{n=0}^N \frac{\sigma_j^n}{n!}} + \sum_j \abs*{\upe^{-\sigma_j} - \sum_{n=0}^N \frac{(-\sigma_j)^n}{n!}}}{\sum_j \abs*{\upe^{\sigma_j} - 1} + \sum_j \abs*{\upe^{-\sigma_j} - 1}} \,.
\end{equation}
Note that ${e_N(\sigma) = \upe^{\sigma} - \sum_{n=0}^N \sigma^n/n! \geq 0}$ because $\sigma \ge 0$. From ${e_0(-\sigma) \leq 0}$, ${e_N(0) = 0}$ and ${\partial_\sigma e_N(-\sigma) = -e_{N-1}(-\sigma)}$ it follows ${e_N(-\sigma) \leq 0}$ for even N and ${e_N(\sigma) \geq 0}$ for odd N, which can be used to eliminate the absolute-value function. Therefore
\begin{equation}
   \frac{\norm{\bm{\varGamma} - \bm{\varGamma}_N}_{\mathrm{Tr}}}{\norm{\bm{\varGamma}}_{\mathrm{Tr}}} =
    \left\{\begin{aligned}
        &\frac{\sum_j [\sinh(\sigma_j) - \mathfrak{s}_N(\sigma_j)]}{\sum_j \sinh(\sigma_j)} & &\text{for even $N$,}\\
        &\frac{\sum_j [\cosh(\sigma_j) - \mathfrak{c}_N(\sigma_j)]}{\sum_j \sinh(\sigma_j)} & &\text{for odd $N$,}
    \end{aligned}\right.
\end{equation}
with $\mathfrak{s}_N$ and $\mathfrak{c}_N$ defined in \cref{eq:truncated_sinh_cosh_definiton}. Applying the mediant inequality yields the desired result in \cref{eq:RenormalizedCovarianceApproximation_RelativeError}.

The application of the mediant inequality is possible because ${f_N(\sigma) = [\sinh(\sigma) - \mathfrak{s}_N(\sigma)]/\sinh(\sigma)}$ and ${h_N(\sigma) = [\cosh(\sigma) - \mathfrak{c}_N(\sigma)]/\sinh(\sigma)}$ increase monotonically:
\begin{IEEEeqnarray}{l}\IEEElabel{eq:RelativeError_Monotonic1}
\sinh^2(\sigma) \diffp{f_N(\sigma)}{\sigma} = \cosh(\sigma) \mathfrak{s}_N(\sigma) - \sinh(\sigma)\cosh_{N-1}(\sigma) \IEEEnonumber\\
= \sum_{n=\frac{N}{2}}^\infty \sum_{m=0}^{\frac{N}{2}-1} \frac{\sigma^{2(n+m)+1}}{(2n)!(2m)!} \mleft( \frac{1}{2m+1} - \frac{1}{2n+1} \mright) \geq 0 \IEEEeqnarraynumspace
\end{IEEEeqnarray}
and
\begin{IEEEeqnarray}{l}\IEEElabel{eq:RelativeError_Monotonic2}
\sinh^2(\sigma) \diffp{h_N(\sigma)}{\sigma} = \cosh(\sigma) \mathfrak{c}_N(\sigma) - \sinh(\sigma) \mathfrak{s}_{N-1}(\sigma) - 1 \IEEEnonumber\\
=  \sum_{n=\frac{N-1}{2}}^\infty  \sum_{m=0}^\infty \frac{\sigma^{2(m+n+1)}}{(2m)!(2n+1)!}\mleft(\frac{1}{2m+1}-\frac{1}{2n+2}\mright) \IEEEnonumber\\
= \sum_{k=0}^\infty \frac{\sigma^{2 k + N + 1}}{(2k+N+1)!} \sum_{l=0}^k \mleft[ \binom{2k+N+1}{2l+1}-\binom{2k+N+1}{2l} \mright]  \IEEEnonumber\\
= \sum_{k=0}^\infty \frac{\sigma^{2 k + N + 1}}{2k + N + 1}\binom{2k+N}{2k+1} \ge 0 \,. \IEEEeqnarraynumspace
\end{IEEEeqnarray}
In \cref{eq:RelativeError_Monotonic1} it was used that all terms for $n < N/2$ vanish due to symmetry and in \cref{eq:RelativeError_Monotonic2} the inner sum was evaluated by employing Egorychev's method~\cite{Riedel_2023,Egorychev_1984}.

\section{Error Bound for the Fredholm Determinant Approximation}\label{sec:Appendix_FredholmDeterminantApproximation}
Suppose we initially have a state described by its renormalized covariance $\bm{\varGamma}$ with eigenvalues $\{\Lambda_j \}$. The state is subject to only passive and loss transformations, before a detection is performed. Then, the error introduced by truncating the Fredholm determinant expansion at order $N$ in \cref{eq:FredholmDeterminant_LogarithmExpansion} is bounded by the eigenvalues of the initial renormalized covariance. Passive transformations preserve the eigenvalues due to their unitarity. As discussed in \cref{sec:Appendix_Gaussian_Transformations}, loss transformations can be described by a beam splitting operation with an auxiliary vacuum mode that is deleted  from the renormalized covariance afterwards. For the sake of this argument, however, we split the loss transformations into two contributions. Let $\eta^2$ be the maximum transmission over all detected modes, that is, the transmission \textit{every} mode reaching one of the detectors of interest experiences when traversing the setup. Then, according to \cref{eq:LossTransformation} this leads to a multiplication of the renormalized covariance with a scalar factor of ${\eta^2 \leq 1}$. The additional losses that some modes experience are treated as passive beam splitter transformations redirecting photons into auxiliary modes, which are later deleted by the projection onto the modes of interest. Thereby, the eigenvalues of the renormalized covariance after traversing the whole setup, before the projection, are given by ${\{ \eta^2 \Lambda_j \}}_j$. It remains to be discussed how the orthogonal projection onto the modes of interest affects the eigenvalues. For the discrete case, where the renormalized covariance is a Hermitian matrix of finite dimension, the answer is given by repeated application of Cauchy's interlacing theorem~\cite{Hwang_2004}. However, in the continuous case, an orthogonal projection of the shape given by \cref{eq:ProjectionContinuum} does not necessarily lower the corresponding space's dimension, so that Cauchy's interlacing theorem cannot be applied directly. However, a similar interlacing inequality holds in the infinite-dimensional case, stating that the relation between the initial eigenvalues $\Lambda_j$ and the eigenvalues after applying an orthogonal projection $\Lambda_j'$ are connected by the relation~\cite{Massey_2023}
\begin{equation}\label{eq:Interlacing}
    \Lambda_{-j} \leq \Lambda_{-j}' \leq 0 \leq \Lambda_j' \leq \Lambda_j \,.
\end{equation}
Using ${\Tr[(-\bm{\varGamma}')^n] = \sum_j ( -\eta^2 \Lambda_j' )^n}$, where $\bm{\varGamma}'$ is the renormalized covariance after the projection, the relative error of the vacuum detection probability
\begin{equation}\label{eq:RelativeErrorVacuumProbability_Definition}
    \frac{\Updelta P_{\vac}^{(N)}}{P_{\vac}} = \frac{\abs*{P_{\vac} - \exp\mleft( \sum_{n=1}^N \frac{\Tr[(-\bm{\varGamma})^n]}{2n}\mright)}}{P_{\vac}}
\end{equation}
can be written as
\begin{equation}
    \frac{\Updelta P_{\vac}^{(N)}}{P_{\vac}} = \mleft| 1 - \exp\mleft( \frac{1}{2} \sum_j \Err_N(\eta^2 \Lambda_j') \mright) \mright| \,,
\end{equation}
with the truncation error of the logarithm given by
\begin{equation}
    \Err_N(\Lambda) = \ln(1+\Lambda) + \sum_{n=1}^N \frac{(-\Lambda)^n}{n} \,.
\end{equation}
Taking the absolute value of the exponent, using the triangle inequality and ${1-e^{-|x|} \leq e^{|x|} - 1}$  yields
\begin{equation}
    \frac{\Updelta P_{\vac}^{(N)}}{P_{\vac}} \leq \exp\mleft( \frac{1}{2} \sum_j | \Err_N(\eta^2\Lambda_j') | \mright) - 1 \,.
\end{equation}
Using \cref{eq:Interlacing} and the fact that $\abs{\Err_N(\pm \Lambda)}$ is monotonically increasing in ${\Lambda > 0}$, the relative error can be connected to the initial eigenvalues $\{ \Lambda_j \}_j$ according to
\begin{equation}
    \frac{\Updelta P_{\vac}^{(N)}}{P_{\vac}} \leq \exp\mleft( \frac{1}{2} \sum_j \abs{ \Err_N(\eta^2 \Lambda_j) } \mright) - 1 \,.
\end{equation}
Thus, the eigenvalues of the initial renormalized covariance bound the relative error in the detection probability when truncating the Fredholm determinant expansion.\\
In some cases, for example when considering highly entangled biphoton states, it is numerically expensive to compute all contributing eigenvalues. For such situations, the triangle inequality can be used again:
\begin{equation}
    \abs{\Err_N(\eta^2 \Lambda_j)} \leq \sum_{\mathclap{n=N+1}}^\infty \frac{\abs{\eta^2 \Lambda_j}^n}{n} \leq \eta^4 \Lambda_j^2 \sum_{\mathclap{n=N+1}}^\infty \frac{|\eta^2 \Lambda_1|^{n-2}}{n} \,,
\end{equation}
where
\begin{equation}
    \sum_{\mathclap{n=N+1}}^\infty \frac{|\eta^2 \Lambda_1|^{n-2}}{n} = -\frac{\ln(1- \eta^2 |\Lambda_1|) + \sum_{n=1}^N \frac{|\eta^2 \Lambda_1|^n}{n}}{\eta^4 \Lambda_1^2} \,.
\end{equation}
Using the fact that the Hilbert-Schmidt norm of the initial renormalized covariance can be expressed as ${\norm{\bm{\varGamma}}_{\mathrm{HS}}^2 = \sum_j \Lambda_j^2}$, this yields
\begin{equation}\label{eq:FredholmDeterminantExpansion_RelativeError}
    \frac{\Updelta P_{\vac}^{(N)}}{P_{\vac}} \leq \mleft( \frac{\exp\mleft( - \sum_{n=1}^N \frac{|\eta^2 \Lambda_1|^n}{n} \mright)}{1 - \eta^2 |\Lambda_1|} \mright)^{\norm{\bm{\varGamma}}_{\mathrm{HS}}^2/(2 \Lambda_1^2)} - 1 \,.
\end{equation}
Thereby, the relative error is bounded in terms of the Hilbert-Schmidt norm and the eigenvalue of largest absolut value $\Lambda_1$ of the initial renormalized covariance.

In many applications, active transformations such as photon-pair generation followed by pump filtering are performed at an early stage while the remainder of the setup employs passive transformations and subsequent detections over a subset of the modes. Thus, we can bound the error of this approximation by constructing the renormalized covariance after the last active transformation and computing only the eigenvalue of largest magnitude as well as the Hilbert-Schmidt norm, either analytically from an appropriate model or numerically by employing a truncated eigenvalue solver.

In case the initial renormalized covariance corresponds to that of a biphoton state, the eigenvalues are completely determined by the Schmidt coefficients of the JSA, obtained by performing a singular value decomposition after introducing a sufficiently fine discretization~\cite{Lamata_2005}. This becomes increasingly expensive to compute for highly entangled states. However, there are algorithms for calculating truncated SVDs efficiently~\cite{Golub_1965,Hochstenbach_2001,Halko_2011,Wu_2015}, which allow to obtain the largest singular values of the JSA and hence the largest eigenvalues of the initial renormalized covariance, without ever performing a full Schmidt decomposition. 

\section*{Acknowledgements}
This research has been funded by the Deutsche Forschungsgemeinschaft (DFG, German Research Foundation), under Grant No.\ SFB 1119--236615297. We thank Robin Krebs for valuable discussions about the complex-valued continuous-mode covariance formalism and the Poisson approximation.

\bibliography{bibliography}% Produces the bibliography via BibTeX.
\end{document}